# Coarse-grained *versus* fully atomistic machine learning for zeolitic imidazolate frameworks

Zoé Faure Beaulieu,[a] Thomas C. Nicholas,[a] John L. A. Gardner,[a] Andrew L. Goodwin,[a,*] and Volker L. Deringer[a,*]

[a] Department of Chemistry, Inorganic Chemistry Laboratory, University of Oxford, Oxford OX1 3QR, UK
* E-mail: andrew.goodwin@chem.ox.ac.uk; volker.deringer@chem.ox.ac.uk

**Zeolitic imidazolate frameworks are widely thought of as being analogous to inorganic AB$_2$ phases. We test the validity of this assumption by comparing simplified and fully atomistic machine-learning models for local environments in ZIFs. Our work addresses the central question to what extent chemical information can be "coarse-grained" in hybrid framework materials.**

Zeolitic imidazolate framework (ZIF) materials[1–4] have garnered interest because of their fundamental properties[5,6] as well as emerging applications.[7,8] ZIFs are a class of metal–organic frameworks (MOFs) with zeolite-like architectures, showing characteristic properties of both groups. Like inorganic zeolites, ZIFs are chemically and thermally stable whilst having markedly higher surface areas and pore volumes. Beyond the crystalline state, ZIFs have been synthesised and characterised in various glassy[9–12] and liquid forms;[13] for a recent review, see Ref. 14.

ZIFs are built up of cationic metal centres and anionic linker molecules. Based on topology and geometry, as well as formal charges (2+/1– on the cations/anions, respectively), ZIFs have long been thought of as tetrahedral AB$_2$ networks analogous to SiO$_2$ (4+/2–; Fig. 1a).[1] Consequently, the conceptual mapping to zeolites has informed the synthesis and understanding of ZIFs.[15,16] However, the extent to which this analogy holds *quantitatively* remains an open question – it is yet unclear whether the energetic landscape of ZIFs can be quantified without a fully atomistic description, and whether the established stability trends in zeolites[17] map onto hybrid ZIF phases, especially given that both materials classes access different crystal topologies.

Describing ZIFs as AB$_2$ networks, as illustrated in Fig. 1b, is an example of structural coarse-graining (cg): a group of atoms or an entire molecule is represented by a single pseudo-atom ("bead"). This approach is popular in computational chemistry, as lowering the structural resolution enables faster simulations; for example, cg dynamics are now abundantly used in biomolecular modelling.[18,19] Conceptually similar cg approaches can help to rationalise interactions in complex disordered networks, as we have recently shown for amorphous calcium carbonate.[20]

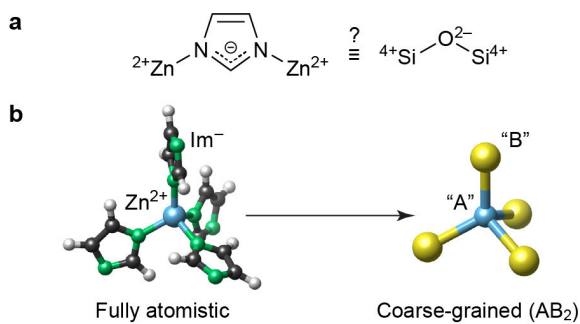

**Fig. 1.** Coarse-graining the structure of the chemically simplest ZIF material, Zn(Im)$_2$, where Im = imidazolate. (**a**) The central question in the present work: to what extent are ZIFs (*left*) similar to silica (*right*), beyond the topological and geometric analogy noted in Ref. 1? (**b**) Schematic of the coarse-graining approach, visualised using VESTA.[21]

In the present work, we test how well ZIFs can be described as coarse-grained AB$_2$ networks from the viewpoint of chemical machine learning (ML). We create ML models for the energies of local environments in ZIFs and compare the accuracy that can be reached with cg *versus* fully atomistic representations. We have previously shown that cg-ML models enable unsupervised learning in this domain, by visualising structural relationships between ZIFs and inorganic AB$_2$ networks.[22,23] Our present work now shows that energetics in ZIFs can be described, to useful accuracy, by supervised cg-ML models. Our study complements wider-ranging activities on cg force-field development,[24–30] and at the same time it addresses general questions about the nature of hybrid framework materials.

To create the input data for this study, we took a large set of AB$_2$-connected networks – which included the experimental ZIF topologies – from Ref. 23. We decorated them with Zn$^{2+}$ cations and Im$^-$ linkers throughout, generated copies of the resulting structures with random distortions, and then evaluated their energies with an empirical force-field model from Ref. 31.‡ In addition to per-cell (*total*) energies, this model by construction yields *local* energies for each atom – allowing us to build a large "synthetic" dataset with which the properties of ML models can be studied, following ideas in Refs. 32 and 33.

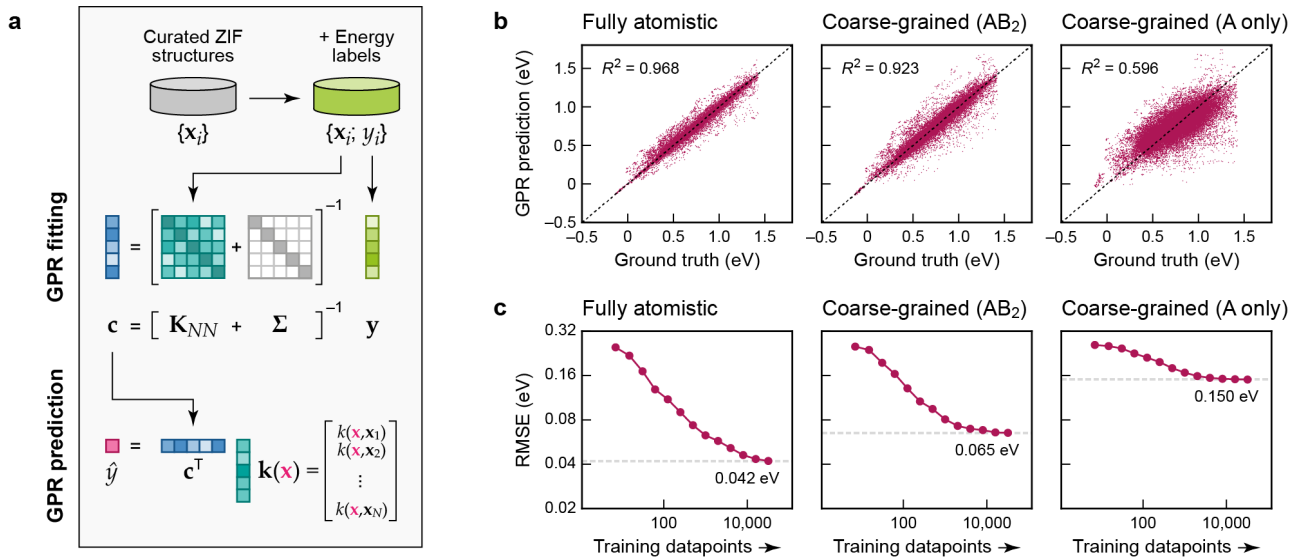

**Fig. 2.** Machine-learning models for the energetic stability of local environments in ZIFs. (**a**) Schematic of Gaussian process regression (GPR), adapted from Ref. 34 and originally published under a CC BY licence (https://creativecommons.org/licenses/by/4.0/). (**b**) Scatter plots of local-environment energies as defined in the text (the "ground-truth" learning target) on the horizontal axis, and our GPR ML predictions on the vertical axis. The values were obtained by 5-fold cross-validation. From left to right, we characterise GPR models based on: a fully atomistic description; a cg description where the linker molecules are described by single "B" beads (Fig. 1b); and a more aggressively coarse-grained model where only A-site species are represented. (**c**) Learning curves showing the model root mean square error (RMSE) depending on the number of training datapoints. The RMSE for the largest number of training points is indicated by a dashed grey line in each panel.

We focus on machine-learning the energetic stability of the $Zn^{2+}$ cations in ZIFs within the present work. We assume that the energetics of $Zn^{2+}$ sites are described by the atomic energies of the cations themselves, $\varepsilon_{Zn}^{(i)}$, and those of their immediate surroundings. Specifically, we assign the energy of any $Im^-$ linker to each of its two $Zn^{2+}$ neighbours in equal parts – just like in $SiO_2$, an O atom that connects two corner-sharing tetrahedra would be attributed half-and-half to both. We obtain the energy of the $j$-th linker, $\varepsilon_{Im}^{(j)}$, by summing over the local energies of all C, N, and H atoms in this particular molecule, and hence,

$$\varepsilon_{local}^{(i)} = \varepsilon_{Zn}^{(i)} + \frac{1}{2}\sum_{j=1}^{4}\varepsilon_{Im}^{(j)}$$

yields the local-environment energy of the $i$-th $Zn^{2+}$ cation in a given ZIF structure, which is our learning target. We note that using the above definition, summing up $\varepsilon_{local}^{(i)}$ over the unit cell conveniently yields the energy of the entire structure, and that the decomposition into local contributions is analogous to a key assumption in ML interatomic potential development.[35,36] We reference the energy values to crystalline ZIF-**zni**, which is the thermodynamically stable form at ambient conditions.[37]

To create atomistic and cg models, we use Gaussian process regression (GPR),[34] an established and data-efficient ML approach. We use the Smooth Overlap of Atomic Positions (SOAP) technique[38] to construct descriptor (feature) vectors, $x_1$ to $x_N$, representing the $Zn^{2+}$ environments, and from those we build the kernel matrix, $K_{NN}$, measuring the similarity of environments with one another. The "ground-truth" labels, $y_1$ to $y_N$, as defined above, are collected in a vector, $y$, and hence the coefficients, $c$, are obtained by solving

$$c = [K_{NN} + \Sigma]^{-1}\,y$$

(Fig. 2a). In this, $\Sigma$ adds a regularisation term, corresponding to expected "noise" in the input, here applied as a constant value for all atoms. Predictions for a new $Zn^{2+}$ environment, $\hat{y}(x)$, are made by computing the SOAP similarity, $k$, between $x$ and every training point, and then evaluating $\hat{y}(x) = c^T k$.‡

We show results for the regression of the local-environment energies in our ZIF dataset in Fig. 2b. The panels in this figure allow us to assess the quality of the GPR ML models compared to the ground-truth (training) data: the scatter plots show how far each prediction deviates from the identity, and therefore how large the numerical error is. The plots show cross-validation results, so that the testing data are not included when training any one specific model.

Moving from the left- to the right-hand side of Fig. 2b allows us to directly compare and contrast a fully atomistic GPR model and the corresponding cg-GPR fits. Even though we have made the structural representation notably simpler, there is only modest loss in accuracy in an $AB_2$-based cg model that includes the locations of the $Im^-$ linkers. In contrast, a purely A-site-based cg model (that represents the ZIF structures by considering only the cation sites and ignoring the location of the linkers) leads to a much poorer prediction of the local-environment energies, because too much structural information has been lost.

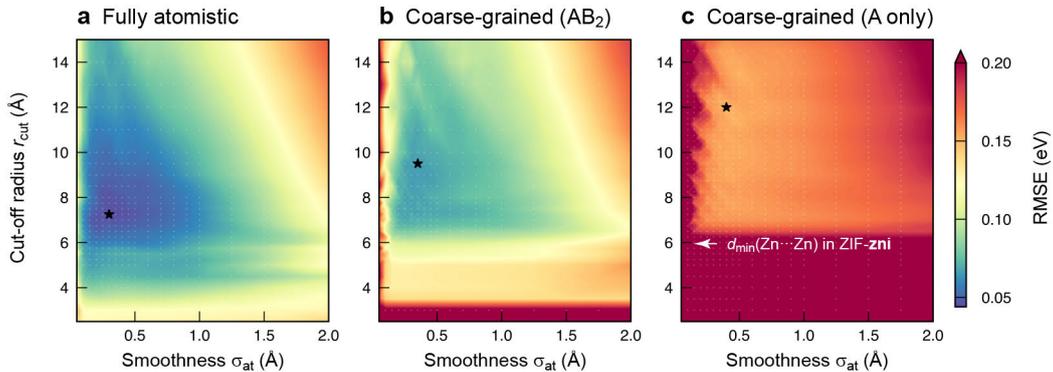

**Fig. 3.** Survey of the hyperparameter space for fully atomistic *versus* cg-GPR models. The two decisive choices are the cut-off radius (vertical axis) and the smoothness of the atomic neighbour density (horizontal axis). The results of a grid search are given by colour-coding, with individual grid points highlighted by small white markers, and the location of the respective minimum indicated by an asterisk. For those scans, we used a more economical setting of $N$ = 20,000 training points, compared to $N$ = 32,000 otherwise.

**Table 1.** RMSE for different GPR models for local-environment energies in ZIFs. The columns show results for the three different representations. The rows correspond to hyperparameters, $\mathcal{H}$, optimised for the respective representation (asterisks in Fig. 3).

| | RMSE for $\varepsilon_{local}^{(i)}$ (eV) | | |
|---|---|---|---|
| | Fully atomistic | cg (AB$_2$) | cg (A only) |
| $\mathcal{H}$ for atomistic | **0.042** | 0.072 | 0.157 |
| $\mathcal{H}$ for cg (AB$_2$) | 0.051 | **0.065** | 0.159 |
| $\mathcal{H}$ for cg (A) | 0.074 | 0.083 | **0.150** |

The learning curves in Fig. 2c, showing the evolution of the error with increasing number of training datapoints, suggest that the ML model predictions are close to converged with regard to the amount of training data used in the fit, particularly in the cg cases. For the fully atomistic GPR model, the quality of prediction is about 40 meV per formula unit, or ≈ 1 kcal mol$^{-1}$. The accuracy decreases by a factor of 1.5 compared to the fully atomistic models if the Im$^-$ units are coarse-grained, but by a factor of more than 3 if they are omitted entirely.

With an initial proof-of-concept in hand, we probed the ML models in more detail by investigating how they depend on the hyperparameters used in the fit. The most important ones control the behaviour of the SOAP descriptor: the cut-off radius, $r_{cut}$, and the smoothness of the atomic neighbour density, $\sigma_{at}$. The former determines the locality of information that the descriptor incorporates for a given environment; the latter affects the ability of the model to generalise to new structures by gradually reducing the precision of the information about atomic sites. Whilst the results in Fig. 2 have been obtained with optimised hyperparameters for each model type, Fig. 3 now shows the results of a grid search that more systematically explores the effect of both hyperparameters on the model error. We sampled values of $r_{cut}$ and $\sigma_{at}$ up to 15 Å and 2 Å, respectively, and thus a much wider range than used in typical SOAP-based ML interatomic potentials ($r_{cut}$ ≈ 5 Å and $\sigma_{at}$ ≈ 0.5 Å; Ref. 34). In each panel of Fig. 3, the respective "best" setting (leading to the lowest numerical error) is indicated by an asterisk.

The loss landscape, *i.e.*, the variation of errors with hyperparameter settings, is quite shallow around the minima, especially in the $r_{cut}$ dimension. Hence, a small change in $r_{cut}$ has little impact on the performance of the model, particularly for the cg ones. The fully atomistic GPR model is optimised by a lower $r_{cut}$ value than its cg counterparts: at a fixed cut-off, the atomistic representation contains more information about the local environment, presumably reducing the number of neighbours (within $r_{cut}$) that need to be included. Similarly, a higher $\sigma_{at}$ is beneficial for the cg models. We cross-checked how using the optimised hyperparameters for one representation affects the prediction error of another (Table 1); the effect is notable for fully atomistic models, much less so for the cg ones.

Comparing the three plots in Fig. 3 side-by-side, it becomes apparent once more how the quality of the models depends on the degree of structural coarse-graining. Reducing the Im$^-$ linkers to single beads leads to models with reasonable errors (Fig. 3b) – whilst leaving them out entirely is clearly unfavourable (Fig. 3c). We also checked the RMSEs for predicting only the Zn$^{2+}$ energies, $\varepsilon_{Zn}^{(i)}$, without taking the energy contributions of the linkers into account; those errors are 0.010, 0.020, and 0.081 eV for the fully atomistic, AB$_2$-, and A-only cg models, respectively, consistent with the trends observed for $\varepsilon_{local}^{(i)}$ (Fig. 2c; Table 1). Future work on ML models for ZIFs might therefore involve fully atomistic and AB$_2$-like cg representations, with the latter reducing the number of coordinates to be considered from 51 per Zn(Im)$_2$ unit to only 9 per AB$_2$ equivalent.

In conclusion, our work has shown that local-environment energies in ZIFs can be "machine-learned" using cg structural representations, with less than a factor of 2 loss of accuracy compared to established, fully atomistic approaches. In doing so, we showed that local energies from an empirical force field for ZIFs[31] can be readily available "synthetic" regression targets – extending prior work in the field of atomistic ML[32,33] to the construction of cg models.

Chemically, our results provide direct and quantitative support for the long-standing idea that there exists a mapping between ZIFs and zeolites (Fig. 1a) based on their underlying tetrahedral connectivity.

What next? A direct avenue for future work is to move from the unsubstituted imidazolate to a wider range of increasingly anisotropic linkers (methyl-, ethyl-, benzimidazolate, *etc.*), and from there onwards to the more general class of AB$_2$ MOFs. In the present proof-of-concept, we focused on the regression of easily available synthetic energies for local Zn$^{2+}$ environments, but we note that GPR and related techniques could similarly be applied to "learning" other atomistic properties, such as NMR chemical shifts.[39–41] Having fast and accurate ML models for predicting the latter could assist in interpreting NMR studies particularly of glassy ZIFs.[42,43] Finally, a clear direction for future research will be the extension from local-energy models to the prediction of forces on atoms, and to the development of cg-ML force fields enabling accurate large-scale simulations for the growing material class of ZIFs.

## Acknowledgements

T.C.N. and J.L.A.G. were supported through an Engineering and Physical Sciences Research Council DTP award [grant number EP/T517811/1]. J.L.A.G. acknowledges a UKRI Linacre - The EPA Cephalosporin Scholarship and support from the Department of Chemistry, University of Oxford. A.L.G. gratefully acknowledges financial support from the E.R.C. (Grant 788144).

## Conflicts of interest

There are no conflicts to declare.

## Notes and references

‡ Data and Python code to reproduce the results are available at https://github.com/ZoeFaureBeaulieu/cg-gpr. The hypothetical ZIF dataset, including technical details of how it was constructed, is available at https://github.com/tcnicholas/hZIF-data.

1. K. S. Park, Z. Ni, A. P. Cote, J. Y. Choi, R. Huang, F. J. Uribe-Romo, H. K. Chae, M. O'Keeffe and O. M. Yaghi, *Proc. Natl. Acad. Sci. U. S. A.*, 2006, **103**, 10186–10191.
2. X.-C. Huang, Y.-Y. Lin, J.-P. Zhang and X.-M. Chen, *Angew. Chem. Int. Ed.*, 2006, **45**, 1557–1559.
3. Y.-Q. Tian, Y.-M. Zhao, Z.-X. Chen, G.-N. Zhang, L.-H. Weng and D.-Y. Zhao, *Chem. Eur. J.*, 2007, **13**, 4146–4154.
4. H. Hayashi, A. P. Côté, H. Furukawa, M. O'Keeffe and O. M. Yaghi, *Nat. Mater.*, 2007, **6**, 501–506.
5. M. T. Wharmby, S. Henke, T. D. Bennett, S. R. Bajpe, I. Schwedler, S. P. Thompson, F. Gozzo, P. Simoncic, C. Mellot-Draznieks, H. Tao, Y. Yue and A. K. Cheetham, *Angew. Chem. Int. Ed.*, 2015, **54**, 6447–6451.
6. S. Henke, M. T. Wharmby, G. Kieslich, I. Hante, A. Schneemann, Y. Wu, D. Daisenberger and A. K. Cheetham, *Chem. Sci.*, 2018, **9**, 1654–1660.
7. R. Banerjee, A. Phan, B. Wang, C. Knobler, H. Furukawa, M. O'Keeffe and O. M. Yaghi, *Science*, 2008, **319**, 939–943.
8. Y. V. Kaneti, S. Dutta, Md. S. A. Hossain, M. J. A. Shiddiky, K.-L. Tung, F.-K. Shieh, C.-K. Tsung, K. C.-W. Wu and Y. Yamauchi, *Adv. Mater.*, 2017, **29**, 1700213.
9. T. D. Bennett, A. L. Goodwin, M. T. Dove, D. A. Keen, M. G. Tucker, E. R. Barney, A. K. Soper, E. G. Bithell, J.-C. Tan and A. K. Cheetham, *Phys. Rev. Lett.*, 2010, **104**, 115503.
10. T. D. Bennett, J.-C. Tan, Y. Yue, E. Baxter, C. Ducati, N. J. Terrill, H. H.-M. Yeung, Z. Zhou, W. Chen, S. Henke, A. K. Cheetham and G. N. Greaves, *Nat. Commun.*, 2015, **6**, 8079.
11. T. D. Bennett, Y. Yue, P. Li, A. Qiao, H. Tao, N. G. Greaves, T. Richards, G. I. Lampronti, S. A. T. Redfern, F. Blanc, O. K. Farha, J. T. Hupp, A. K. Cheetham and D. A. Keen, *J. Am. Chem. Soc.*, 2016, **138**, 3484–3492.
12. C. Zhou, L. Longley, A. Krajnc, G. J. Smales, A. Qiao, I. Erucar, C. M. Doherty, A. W. Thornton, A. J. Hill, C. W. Ashling, O. T. Qazvini, S. J. Lee, P. A. Chater, N. J. Terrill, A. J. Smith, Y. Yue, G. Mali, D. A. Keen, S. G. Telfer and T. D. Bennett, *Nat. Commun.*, 2018, **9**, 5042.
13. R. Gaillac, P. Pullumbi, K. A. Beyer, K. W. Chapman, D. A. Keen, T. D. Bennett and F.-X. Coudert, *Nat. Mater.*, 2017, **16**, 1149–1154.
14. T. D. Bennett and S. Horike, *Nat. Rev. Mater.*, 2018, **3**, 431–440.
15. A. Phan, C. J. Doonan, F. J. Uribe-Romo, C. B. Knobler, M. O'Keeffe and O. M. Yaghi, *Acc. Chem. Res.*, 2010, **43**, 58–67.
16. B. Chen, Z. Yang, Y. Zhu and Y. Xia, *J. Mater. Chem. A*, 2014, **2**, 16811–16831.
17. A. Sartbaeva, S. A. Wells, M. M. J. Treacy and M. F. Thorpe, *Nat. Mater.*, 2006, **5**, 962–965.
18. S. Riniker, J. R. Allison and W. F. van Gunsteren, *Phys. Chem. Chem. Phys.*, 2012, **14**, 12423–12430.
19. W. G. Noid, *J. Chem. Phys.*, 2013, **139**, 090901.
20. T. C. Nicholas, A. E. Stones, A. Patel, F. M. Michel, R. J. Reeder, D. G. A. L. Aarts, V. L. Deringer and A. L. Goodwin, *arXiv preprint*, 2023, arXiv:2303.06178 [cond-mat.mtrl-sci].
21. K. Momma and F. Izumi, *J. Appl. Crystallogr.*, 2011, **44**, 1272–1276.
22. T. C. Nicholas, A. L. Goodwin and V. L. Deringer, *Chem. Sci.*, 2020, **11**, 12580–12587.
23. T. C. Nicholas, E. V. Alexandrov, V. A. Blatov, A. P. Shevchenko, D. M. Proserpio, A. L. Goodwin and V. L. Deringer, *Chem. Mater.*, 2021, **33**, 8289–8300.
24. S. T. John and G. Csányi, *J. Phys. Chem. B*, 2017, **121**, 10934–10949.
25. T. Lemke and C. Peter, *J. Chem. Theory Comput.*, 2017, **13**, 6213–6221.
26. L. Zhang, J. Han, H. Wang, R. Car and W. E, *J. Chem. Phys.*, 2018, **149**, 034101.
27. K. K. Bejagam, S. Singh, Y. An and S. A. Deshmukh, *J. Phys. Chem. Lett.*, 2018, **9**, 4667–4672.
28. H. Chan, M. J. Cherukara, B. Narayanan, T. D. Loeffler, C. Benmore, S. K. Gray and S. K. R. S. Sankaranarayanan, *Nat. Commun.*, 2019, **10**, 379.
29. J. Wang, S. Olsson, C. Wehmeyer, A. Pérez, N. E. Charron, G. de Fabritiis, F. Noé and C. Clementi, *ACS Cent. Sci.*, 2019, **5**, 755–767.
30. W. Wang and R. Gómez-Bombarelli, *npj Comput. Mater.*, 2019, **5**, 125.
31. J. P. Dürholt, G. Fraux, F.-X. Coudert and R. Schmid, *J. Chem. Theory Comput.*, 2019, **15**, 2420–2432.
32. Z. Shui, D. S. Karls, M. Wen, I. A. Nikiforov, E. Tadmor and G. Karypis, in *Advances in Neural Information Processing Systems*, eds. S. Koyejo, S. Mohamed, A. Agarwal, D. Belgrave, K. Cho and A. Oh, Curran Associates, Inc., 2022, vol. 35, pp. 14839–14851.

33  J. L. A. Gardner, Z. Faure Beaulieu and V. L. Deringer, *Digital Discovery*, 2023, Advance Article, DOI: 10.1039/D2DD00137C.
34  V. L. Deringer, A. P. Bartók, N. Bernstein, D. M. Wilkins, M. Ceriotti and G. Csányi, *Chem. Rev.*, 2021, **121**, 10073–10141.
35  J. Behler and M. Parrinello, *Phys. Rev. Lett.*, 2007, **98**, 146401.
36  A. P. Bartók, M. C. Payne, R. Kondor and G. Csányi, *Phys. Rev. Lett.*, 2010, **104**, 136403.
37  J. T. Hughes, T. D. Bennett, A. K. Cheetham and A. Navrotsky, *J. Am. Chem. Soc.*, 2013, **135**, 598–601.
38  A. P. Bartók, R. Kondor and G. Csányi, *Phys. Rev. B*, 2013, **87**, 184115.
39  F. M. Paruzzo, A. Hofstetter, F. Musil, S. De, M. Ceriotti and L. Emsley, *Nat. Commun.*, 2018, **9**, 4501.
40  Z. Chaker, M. Salanne, J.-M. Delaye and T. Charpentier, *Phys. Chem. Chem. Phys.*, 2019, **21**, 21709–21725.
41  W. Gerrard, L. A. Bratholm, M. J. Packer, A. J. Mulholland, D. R. Glowacki and C. P. Butts, *Chem. Sci.*, 2020, **11**, 508–515.
42  E. F. Baxter, T. D. Bennett, C. Mellot-Draznieks, C. Gervais, F. Blanc and A. K. Cheetham, *Phys. Chem. Chem. Phys.*, 2015, **17**, 25191–25196.
43  R. S. K. Madsen, A. Qiao, J. Sen, I. Hung, K. Chen, Z. Gan, S. Sen and Y. Yue, *Science*, 2020, **367**, 1473–1476.